\documentclass[preprint2]{aastex}
\usepackage{amssymb,amsmath}
\usepackage{epstopdf}
\usepackage{graphicx}
\usepackage{url}
\usepackage{textcomp}

\shorttitle{Detecting Pop III GRB infrared afterglows}
\shortauthors{Macpherson, Coward, and Zadnik}

\begin{document}

\title{The Potential for Detecting Gamma-Ray Burst Afterglows from Population III Stars with the Next Generation of Infrared Telescopes}


\author{D. Macpherson\altaffilmark{1}}
\affil{ICRAR, University of Western Austalia, Crawley WA 6009, Australia}
\email{damien.macpherson@icrar.org}

\and

\author{D.M. Coward\altaffilmark{2}}
\affil{School of Physics, University of Western Australia, Crawley WA 6009, Australia}

\and

\author{M.G. Zadnik}
\affil{Department of Imaging and Applied Physics, Curtin University, Western Australia}


\altaffiltext{1}{School of Physics, University of Western Australia, Crawley WA 6009, Australia} 

\altaffiltext{2}{Australian Research Council Future Fellow}

\begin{abstract}
We investigate the detectability of a proposed population of Gamma-Ray Bursts (GRBs) from the collapse of Population III (Pop III) stars. The James Webb Space Telescope (JWST) and Space Infrared telescope for Cosmology and Astrophysics (SPICA) will be able to observe the late time infrared afterglows. We have developed a new method to calculate their detectability, which takes into account the fundamental initial mass function (IMF) and formation rates of Pop III stars, from which we find the temporal variability of the afterglows and ultimately the length of time JWST and SPICA can detect them. In the range of plausible Pop III GRB parameters, the afterglows are always detectable by these instruments during the isotropic emission, for a minimum of 55 days and a maximum of 3.7 years. The average number of detectable afterglows will be 2.96$\times 10^{-5}$ per SPICA field of view (FOV) and 2.78$\times 10^{-6}$ per JWST FOV. These are lower limits, using a pessimistic estimate of Pop III star formation. An optimal observing strategy with SPICA could identify a candidate orphan afterglow in $\sim 1.3$ years, with a 90 percent probability of confirmation with further detailed observations. A beamed GRB will align with the FOV of the planned GRB detector Energetic X-ray Imaging Survey Telescope once every 9 years. Pop III GRBs will be more easily detected by their isotropic emissions (i.e. orphan afterglows) rather than by their prompt emissions.
\end{abstract}


\section{INTRODUCTION}
Gamma-Ray Bursts (GRBs) are the brightest observable events ever discovered, and are visible from greater distances than any other astronomical phenomenon. Several GRBs have been detected at the end of the Epoch of Reionisation, at redshifts greater than 6 (GRB 050904 at z=6.3 (Tagliaferri et al. 2005), GRB 080913 at z=6.7 (Greiner et al. 2009), GRB 090423 at z=8.2 (Tanvir et al. 2009)). The highest redshift previously measured was z=9.4 for GRB 090429B (Cucchiara et al. 2011). Until recently, this was the most distant object ever detected. It was recently surpassed by MACS 1149-JD, a young, small galaxy at a redshift of 9.6 (Zheng et al. 2012).

The extreme brightness of GRBs makes them ideal probes of the high redshift universe. As such, it is hoped that GRBs will provide a window to observe the universe at earlier times than ever before.

In this paper we investigate the prospects of detecting the afterglows of GRBs from the first stars, so-called Population III (Pop III stars). There has been much theoretical effort on the formation and fate of Pop III stars, which has largely remained untested due to the difficulty in observing them.

The prospects of direct detection hinges upon groups of such stars forming at relatively low redshifts (z$\sim$6), in improbably pristine isolated halos (Johnson, Dalla \& Kochfar 2012), or through gravitational lensing amplifying the light of the distant stars (Zackrisson et al. 2012). The best prospects for detection are from emissions at the end of the stars' lives. Theorists have speculated that, if a Pop III star meets certain evolutionary constraints, it will end as a GRB brighter and more energetic than any burst yet detected (Toma, Sakamoto \& M\'{e}sz\'{a}ros 2011; Yoon, Dierks \& Langer 2012; Nagakura, Suwa \& Ioka 2011; Campisi et al. 2011; M\'{e}sz\'{a}ros \& Rees 2010; Komissarov \& Barkov 2010; Nakauchi et al. 2012). One of these constraints is a high rotation rate which Pop III stars are thought to have based on cosmological simulations (Stacy, Bromm \& Loeb 2011) and chemical signatures of low metallicity stars in the oldest Galactic globular cluster (Chiappini et al. 2011).

Exactly when in the history of the universe we should find Pop III stars is uncertain; currently we have only the results of simulations from various models. Modelled Pop III star formation rates vary, with peak values between 10$^{-4}$ and 10$^{-5}$ M$_{\odot}$Mpc$^{-3}$yr$^{-1}$ between redshifts 7 and 20 (Trenti \& Stiavelli, 2009; Maio et al. 2010; Johnson et al. 2012; Wise et al. 2012). The most recent Pop III stars could potentially form in isolated halos of pure hydrogen at z $\sim$ 6 (Johnson et al. 2012), while the earliest limits on star formation may be around z $\sim$ 60 (Naoz \& Bromberg, 2007). As redshift to the burst increases, the optical afterglow redwards of the Lyman-$\alpha$ line is shifted to longer and longer wavelengths, progressively moving out of optical filter bandpasses. In order to detect the optical afterglow of a GRB at high redshift, we must apply an equivalent redshift to our detectors by moving our focus from optical-NIR to Mid-IR.

The Mid-IR detectors primarily considered here are the planned James Webb Space Telescope (\textit{JWST}) and Space Infrared telescope for Cosmology and Astrophysics (\textit{SPICA}). Both instruments are planned to launch in 2018, and will provide a significant advancement in sensitivity over previous instruments to observe in Mid-IR. Because these instruments are still under construction, with many details yet to be finalised, we must stress that the afterglow lightcurves we determine in this paper are based upon the characteristics of the instruments as they are currently expected or intended to be. In the case of JWST we at least know which filters will be used and their response functions (Glasse et al. 2010), and the target sensitivities of the instrument through those filters \footnote{STScI Prototype JWST Exposure Time Calculator, \url{http://jwstetc.stsci.edu/etc/input/miri/imaging/}}. 

For SPICA the filters have yet to be determined, so we assume three bandpasses distributed across the Mid Infrared Camera and Spectrometer (MCS) spectral range (5$\mu$m to 38$\mu$m) and estimated the instrument's sensitivity in those bands from recent publications of the Wide Field Camera's expected performance (Kataza 2012, \textit{conference presentation}\footnote{\url{http://www.ioa.s.u-tokyo.ac.jp/gopira/sym2012/proceeding/Kataza_v1.ppt}}).

\subsection{A Framework for Estimating a Detection Rate}
For determining the observability of Pop III GRB afterglows we diverge from the typical method of applying a luminosity function to the intrinsic rate of events. Our method is based instead on the intrinsic rate and the length of time the afterglow can be observed.

We construct an energy distribution based on models of the Pop III formation rate, initial mass function (IMF), and GRB energetics. Instead of combining this with afterglow modelling to find the distribution of observed luminosity at a certain time (a luminosity function), we extend the modelling to find the distribution of the limiting time (the time interval between the initial prompt and when the observed flux density diminishes below the 5$\sigma$ limit of the detector).

The key advantage of calculating the distribution of the limiting time is revealing whether the afterglow can be detected after the jet break time. This point is critical for estimating the detection rate, since prior to this time the afterglow emission is beamed within the jet opening angle. An instrument that cannot detect the post-jet break emission is only sensitive to the small percentage of GRBs directed toward it. An instrument that can detect the post-jet break emission is sensitive to the entire intrinsic rate of GRBs.

A second advantage of employing temporal detection durations is determining if there would be an overlap between successive afterglows, where the rate of bursts is such that one appears while the afterglow of the previous is still visible. This affects the number visible at any given time, and therefore also the chances of detection.

By multiplying the limiting time by the intrinsic all-sky rate of events one obtains the instantaneous number per sky area. This is the average number detectable across the whole sky at any given time. This is a different type of rate to that typically calculated using the intrinsic rate, beaming factor, and luminosity function, which is the number of detections for a given instrument over a given time interval. The number density reflects the probability of detection in any given area, and therefore the average amount of imaging required per event. It is best suited to describing the prospects of detecting very rare but long lasting transient events, i.e. Pop III GRB afterglows.

In this paper the modelling is described in Section 2. Section 2.1 deals with the properties of the GRBs expected from Pop III stars. Section 2.2 outlines how we simulate the afterglows of Pop III GRBs, in order to show how we go about determining the limiting time. In Section 2.3 we extend this to find how the limiting time is affected by burst energy and redshift. In Section 3  we combine the intrinsic rates of GRBs with the afterglow limiting time, both as functions of energy and redshift, to then find the instantaneous number density of detectable afterglows. Conclusions are presented in Section 4, and the implications are discussed in Section 5.

\section{MODELLING}
While there are several parameters which impact upon the properties of the GRB afterglow, and thus its limiting time, in this paper we focus on those connected to the initial stellar properties. These properties, specifically the formation rate and mass distribution, should have a measurable impact on the local Universe, and thus can be inferred without observation of the GRBs themselves. For instance, the jet opening angle has a strong influence on the observable characteristics of the GRB and its afterglow, but can only be measured by observing the afterglow. On the other hand, the formation rate and mass distribution will also affect the energy and redshift distributions of the GRBs.

From \citet{desouza2011} we extract the rate of Pop III GRB formation as a function of redshift, in their most pessimistic case (see their figure 5). The stars are divided into two sub-populations based on their formation rates and initial masses. Pop III.1 formed earlier, peaking at $\sim 1.2 \times 10^{-3}$ GRB yr$^{-1}$ at z $\sim 17$, with mass in the range $\sim 100 - 1000M_{\odot}$. Pop III.2 stars formed at a much higher rate under the influence of radiation from Pop III.1 stars, peaking at $\sim 4.6 \times 10^1$ GRB yr$^{-1}$ at z $\sim 9$, with mass in the range $\sim 10 - 100M_{\odot}$.

\begin{figure}
\centering
\plotone{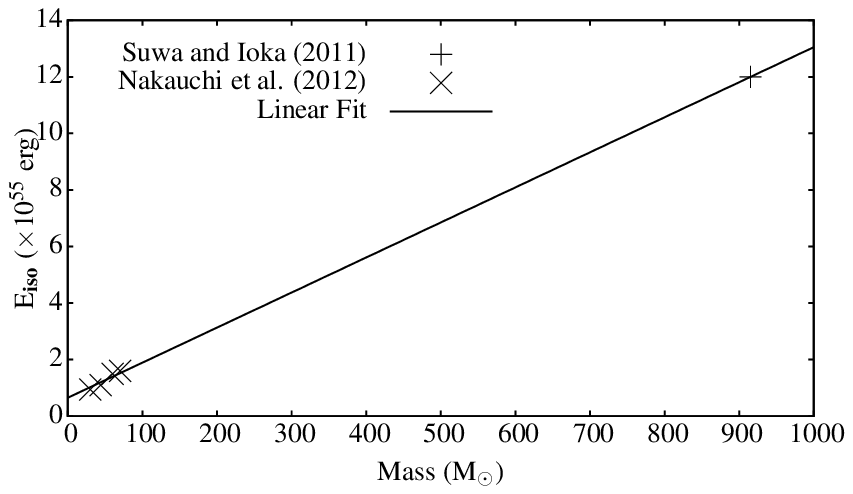}
\caption{\label{fig1}Mass-energy relation extracted from the results of Nakauchi et al. (2012) and Suwa \& Ioka (2011). The linear fit is described by the equation $E_{iso} (\times 10^{55} erg)=0.0124 M(M_{\odot})+0.6507$}
\end{figure}

By comparing the work of \citet{nakauchi2012} and \citet{suwa2011} we found a linear relation between the initial mass of the star and the isotropic equivalent energy ($E_{iso}$) of its resultant GRB (see Figure \ref{fig1}). The linear fit is supported by the presumption that Pop III GRBs are the result of Poynting-flux dominated jets, for which the total energy is roughly proportional to the mass of the progenitor \citep{toma2011}.

If a Pop III.2 star became a red supergiant (RSG), then in the collapsar scenario of GRB formation the accretion of matter onto the nascent black hole would not last long enough to drive a jet through the stellar envelope. If on the other hand the star's final stage were a blue supergiant (BSG), the jet could penetrate the smaller envelope and cause a GRB.  \citet{yoon2012} claimed that a Pop III star would end as a BSG if it could undergo chemically homogeneous evolution, then defined the initial mass and rotation velocity parameter space where this would occur (see their figure 12). They concluded that the mass of the star must be between 13 and 84 $M_{\odot}$. This mass range is supported by the numerical simulations of \citet{nakauchi2012}.

The models and predictions for Pop III.1 GRBs tend to focus on the very top end of the mass distribution. When the mass of the progenitor is near 1000 $M_{\odot}$, it is argued that the mass of the central black hole and surrounding accretion torus would both be of the order $10^2M_{\odot}$. Therefore, the accretion time would be sufficient to power a Poynting-flux dominated jet until it penetrated the stellar envelope, even if the star were an RSG \citep{toma2011,meszaros2010,komissarov2010,nagakura2012,suwa2011}.

We can conclude from these arguments that a significant fraction of Pop III.2 stars are capable of producing GRBs, whereas the only candidates for Pop III.1 GRBs are from the very small proportion of stars at the upper limits of the mass distribution. Given that Pop III.1 star formation is much less than Pop III.2 star formation, the low proportion of sufficiently large progenitors makes the contribution of Pop III.1 stars to the GRB rate negligible. In the following sections we shall focus only on Pop III.2 GRB progenitors in the mass range defined by \citet{yoon2012}, revisiting the very high mass Pop III.1 GRBs only briefly in the discussion.

\subsection{Properties of Pop III Progenitors}
Applying the linear mass-energy relation shown in Figure \ref{fig1} to the GRB progenitor mass range of 13 - 84$M_{\odot}$ \citep{yoon2012} leads to an isotropic equivalent energy range of $\sim 0.8$ to $\sim 1.7 \times 10^{55}$ erg. The distribution of energies within this range is found by combining the mass-dependent rotation limits of the formation region with the IMF.

In performing this combination we have not presumed any distribution in the initial rotation velocity of Pop III.2 stars. We consider that any given star has an equal probability of acquiring any rotation velocity up to the break-up limit. The probability that a star will have the necessary mass and rotation to produce a GRB is then the difference between the minimum rotation for GRB formation ($v_{GRB, min}(m)$) and the break-up rotation ($v_{max}(m)$), divided by the break-up rotation at that mass (${v_{max}(m)}$).

\begin{eqnarray}
\label{eqnPgrb}
P_{GRB}(m)=\frac{v_{max}(m)-v_{GRB, min}(m)}{v_{max}(m)}.
\end{eqnarray}

The probability in equation \ref{eqnPgrb} is conservative, given that Pop III stars are expected to be fast rotators \citep{stacy2011, chiappini2011}. This probability is multiplied by the IMF (Salpeter to be consistent with the pessimistic formation rate from \citet{desouza2011}), the result normalised, and finally converted to a distribution of $E_{iso}$ using the relation shown in Figure \ref{fig1}.

\begin{figure}
\centering
\plotone{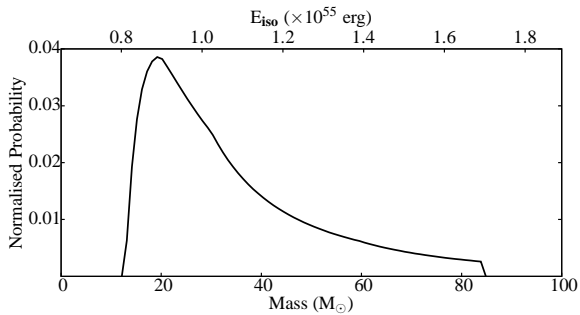}
\caption{\label{fig3x}Normalised mass distribution of Population III.2 GRB progenitors, and the associated $E_{iso}$ distribution of those GRBs.}
\end{figure}

Figure \ref{fig3x} shows the normalised probability distribution of both Pop III GRB progenitor mass and resultant energy. The distribution has a sharp peak at $\sim 20 M_{\odot} (\sim 9 \times 10^{54}\text{erg})$ followed by a slow decay at higher masses. This is similar to the usual Salpeter decay, however incorporating the \citet{yoon2012} GRB formation region sharply reduces the contribution of low mass progenitors.

Since we used different GRB formation constraints, it became necessary to adjust the GRB formation rate accordingly. The ratio of the sizes of the formation regions of \citet{yoon2012} $\left( \int\limits_{13}^{84} \phi(m) P_{GRB}(m) \text{ d}m\right) $ and \citet{desouza2011} $\left( \int\limits_{25}^{100}\phi(m) \text{ d}m\right) $, where $\phi(m)$ is the Salpeter IMF, is 0.51. We have scaled the Pop III.2 GRB formation rate of \citet{desouza2011} in the pessimistic case by this value.

\begin{figure}
\centering
\plotone{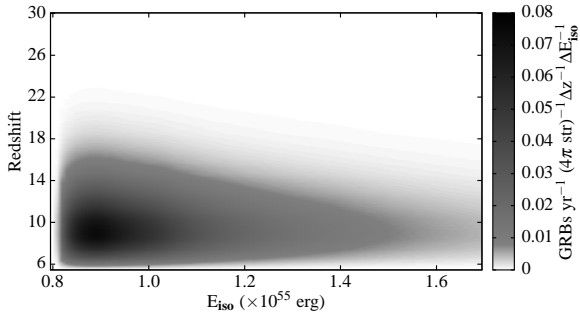}
\caption{\label{fig4}Intrinsic rate of GRBs from Pop III.2 progenitors in terms of redshift and $E_{iso}$ in the most pessimistic case of \citet{desouza2011}, modified by the GRB formation constraints of \citet{yoon2012}}
\end{figure}

A map of the Pop III.2 GRB formation rate in terms of $E_{iso}$ and redshift is shown in Figure \ref{fig4}. The rate of GRBs in each section of the map has been calculated by multiplying the integrated modified differential GRB rate over 0.1 z by the integrated normalised energy distribution (Figure \ref{fig3x}) over $10^{53}$ erg. For example, the modified rate of bursts integrated over $z = 9.95 \text{ to } 10.05$ is 2.21 yr$^{-1}$. The normalised energy probability integrated over $E_{iso} = 9.95 \text{ to } 10.05 \times 10^{54}$ erg is 0.0221. The rate of Pop III GRBs occurring at $z = 10 \pm 0.05$ with $E_{iso} = 10 \pm 0.05 \times 10^{54}$ erg is 0.0488 yr$^{-1}$.

Figure 3 shows that the majority of GRBs are at redshifts $<18$. The rate of bursts from redshifts $\geq$18 account for 2\% of the total rate of bursts from Pop III.2 stars (2.9 bursts yr$^{-1}$). The rate peaks at 7.38$\times 10^{-2}$ yr$^{-1}$ with $E_{iso}=8.9\pm 0.05\times 10^{54}$ erg and $z=9\pm 0.05$. Total rate summed over the entire range of energy and redshift is $1.45\times 10^{2}$ yr$^{-1}$.

\subsection{Afterglow Simulations}
In simulating the lightcurves of the afterglow, we use the algorithm published in the appendix of \citet{toma2011}. This assumes that the afterglow is caused by synchrotron emission when the jet launched from the black hole creates a shock front with the ISM. Throughout this paper we have adopted their fiducial parameters: external medium density $n_0=1$ cm$^{-3}$ (c.f. \citet{chandra2010}), magnetic energy fraction $\epsilon_B=10^{-2}$, electron energy fraction $\epsilon_e=10^{-1}$, electron energy index $p=2.3$, and jet opening angle $\theta_j=10^{-1}$ rad. For more details of the algorithm see their paper.

The final output of their equation is the flux $\varepsilon F_\varepsilon$ as a function of energy and time. We use the following equation to convert this to a flux density $F_\nu$ in a given filter band:

\begin{eqnarray}
F_\nu&=\dfrac{\int \varepsilon F_\varepsilon (h/\varepsilon) S(\varepsilon) d\varepsilon}{\int S(\varepsilon) d\varepsilon},
\end{eqnarray}

where $S(\varepsilon)$ is the filter response function of the filter band in terms of energy.

By taking those fiducial values as given, the only variables left to the algorithm are the isotropic equivalent energy and redshift, which we determined in the previous subsection.

\begin{figure}
\centering
\plotone{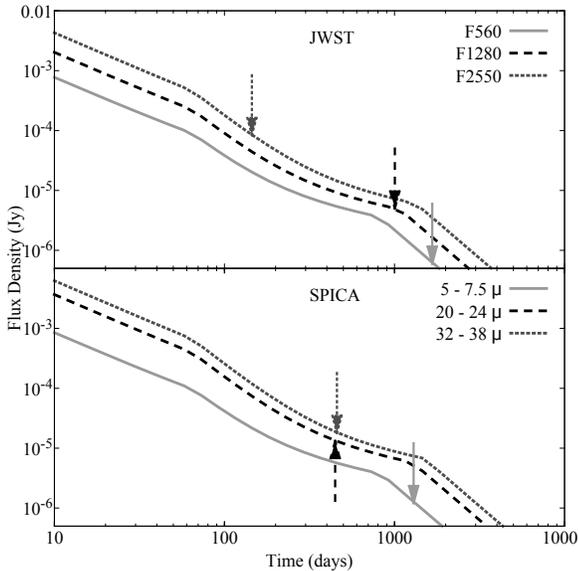}
\caption{\label{fig5}Afterglow light curve of one of the most likely GRBs from a Pop III progenitor ($E_{iso} = 8.9\times 10^{54}$ erg, z = 9) viewed with selected IR filter bands of JWST (\textbf{Top}) and SPICA (\textbf{Bottom}). With these parameters, the afterglow exhibits a jet break at 65 days. The arrows show the limiting times of the afterglow through the various filters, i.e the points where the 5$\sigma$ flux density limit of a 300s exposure with that filter exceeds the flux density of the afterglow. With JWST these limiting times are 1,667 days for F560, 1,003 days for F1280, and 144 days for F2550. With SPICA they are 1,294 days for  5-7.5\micron, 447 days for 20-24\micron, and 458 days for 32-38\micron.}
\end{figure}

Figure \ref{fig4} showed that the highest rate of Pop III bursts are those from a redshift between 8.95 and 9.05, with $E_{iso}$ between 8.85 and 8.95 $\times 10^{54}$ erg (rate of $7.38\times 10^{-2}$ bursts yr$^{-1}$). We have simulated an afterglow using the method of \citet{toma2011} with the parameters $E_{iso} = 8.9\times 10^{54}$ erg and z = 9, as would be viewed with a number of infrared filter bands to be incorporated in the planned JWST and SPICA space telescopes, shown in Figure \ref{fig5}.

The selected filter bands span the extent of the respective instruments' IR coverage. From the positions of the arrows in Figure \ref{fig5}, denoting the limiting times of the afterglow, one can see the effects of the different designs of the telescopes. JWST has better sensitivity than SPICA at wavelengths up to $\sim$ 15 \micron, but at longer wavelengths SPICA is more sensitive.

Figure \ref{fig5} shows that the planned JWST and SPICA instruments will be well suited to observing GRB afterglows at high redshift. In this figure, the second band to lose sight of the afterglow is SPICA 20 - 24\micron. At the time when this band loses the afterglow, every other SPICA band and half of the JWST IR filters can still observe it.

Of greatest significance to the detection of these afterglows is that the jet break occurs earlier than the limiting times for all filters. After the jet break, the radiation is no longer beamed, and can be seen regardless of the orientation of the GRB. Being sensitive to the afterglow while it is emitting isotropically will make the JWST and SPICA telescopes theoretically capable of detecting the intrinsic number of GRB afterglows. This drastically increases the potential number of detections, as with a jet opening angle of 0.1 rad the fraction of the sky which sees the beamed emission is only 0.05\%.

\subsection{Afterglow Duration}
Extending the simulation of the afterglows we find the afterglow durations of GRBs over the range of plausible $E_{iso}$ and z for which the event rates were plotted in Figure \ref{fig4}. We have chosen to find the limiting times with the SPICA 20 - 24\micron$\,$ band, as when this band loses sight of the afterglow, there are still potentially half a dozen other available filters (see Figure \ref{fig5}).

It is important to note at this point that hereafter when we refer to the duration of an afterglow we refer to the isotropic phase, when the emission is not beamed. This is the time interval between the jet break time and the limiting time. In Figure \ref{fig5}, the jet break appears as the achromatic kink in the lightcurves, approximately 65 days after the burst. Given that the afterglow flux density in the SPICA 20 - 24\micron$\,$ band is exceeded by that instrument's 5$\sigma$ limit after 447 days, the total isotropic duration with this filter is 382 days.

Throughout the plausible parameter space of Pop III GRBs, we have calculated the jet break time and limiting time of the afterglow, to find the isotropic duration. The jet break time is found via the equation \citep{toma2011}

\begin{eqnarray}
\left(\dfrac{E_{iso}\theta_j^8}{4\pi n m_p c^5}\right)^{1/3}(1+z).
\end{eqnarray}

\begin{figure}
\centering
\plotone{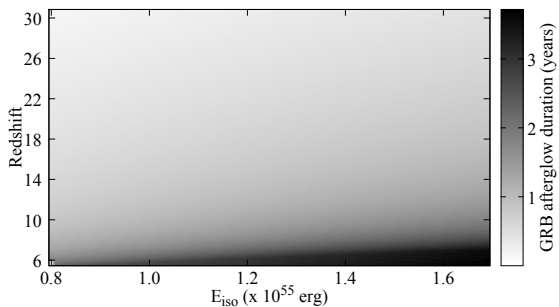}
\caption{\label{fig6}Isotropic afterglow duration of GRBs with parameters spanning Pop III progenitors.}
\end{figure}

If the flux density at this time is less than the limit of the detector, the duration is said to be zero. Otherwise, we start at this time and step forward until the flux density is less than the limit, and the time between this and the jet break time is the duration. This is plotted in Figure \ref{fig6}.

Figure \ref{fig6} also shows the prime reason we change from considering the overall limiting time to the isotropic duration: at all points in the parameter space shown in this figure the isotropic duration is $>0$. Thus as long as our fixed model parameters are accurate and our free parameters remain within the expected ranges, the resultant GRB afterglows will always be visible to JWST and SPICA after the jet break. Thus a strategy to detect orphan afterglows with these instruments would capture the entire intrinsic rate of contributions from Pop III stars.

\section{RESULTS: IR DETECTION}
Here we describe finding the all-sky number density of transient objects with a known duration and rate of occurrence.

The number density is the average detectable number at any given time within a certain area, and reflects the probability of detection within that area. A number density of 1 in some area means that, on average, there will always be a detectable object within that area. This would be caused by the duration being the inverse of the intrinsic rate of events within that area.

Hereafter we calculate the number density of Pop III GRB afterglows, for those that have an observed flux density in the 20 - 24\micron$\,$ band greater than 13.2$\mu$ Jy, which is the SPICA instrument's planned 5$\sigma$ limit for a 300s exposure in this band.

The number density (number of detectable afterglows $4\pi$ str$^{-1}$) is a product of the intrinsic all-sky burst rate (number of bursts $4\pi$ str$^{-1}$ yr$^{-1}$) with the duration (years between the jet break time and the afterglow reaching the limiting flux density). We first find this for each individual subsection of the parameter space of Pop III GRBs. For example, Figure \ref{fig4} shows that the rate of bursts with $E_{iso} = 8.9\pm 0.05\times 10^{54}$ erg at $z = 9\pm 0.05$ is $7.38\times 10^{-2}$ yr$^{-1}$/4$\pi$ str. Figure \ref{fig5} shows that the SPICA 20 - 24\micron$\,$ band can see the afterglow of a burst with $E_{iso} = 8.9\times 10^{54}$ erg at $z = 9$ for 1.05 yr (382 days) after the jet break. Multiplying the rate of bursts yr$^{-1}$/4$\pi$ str by the number of years for which they are visible results in $7.72\times 10^{-2}$ afterglows/4$\pi$ str.

\begin{figure}[h]
\plotone{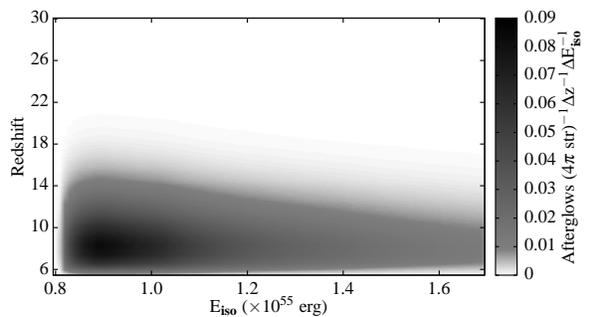}
\caption{\label{fig8}All sky number of visible Pop III.2 GRB afterglows in the plausible parameter space. It is the all sky (4$\pi$ str) number of afterglows whose flux density exceeds the 5$\sigma$ limit of the SPICA instrument in the 20 - 24\micron$\,$ band at any given time.}
\end{figure}

Proceeding in this manner, Figure \ref{fig8} shows a distribution of the all-sky number density of Pop III GRB afterglows as a function of both $E_{iso}$ and z. This figure is effectively Figure \ref{fig4} multiplied by Figure \ref{fig6}.

The number density is diminished towards high redshift and is extended in the energy axis, compared to the plot of the GRB rate (Figure \ref{fig4}). The increase in duration caused by increased energy has not moved the peak contribution from that in Figure \ref{fig4}; it is still $8.9\pm 0.05 \times 10^{54}$ erg. However the reduction in duration caused by increasing redshift has caused the number density redshift peak to move to $8.4 \pm 0.05$.

By summing all number densities of this parameter space, we find that the number density of Pop III.2 GRB afterglows is $1.76\times 10^2/4\pi$str for the SPICA 20 - 24\micron$\,$ band.

\section{CONCLUSIONS}
We derive a linear mass-energy relationship which satisfies the expectation for the energy of a Poynting-flux dominated jet to be approximately proportional to the mass of the progenitor star \citep{nakauchi2012, suwa2011, toma2011}. Using this in combination with model predictions of the mass and rotation characteristics necessary for a Pop III star to produce a GRB \citep{yoon2012}, we find that the minimum isotropic equivalent energy of a Pop III GRB is $\sim 0.8\times 10^{55}$ erg.

Combining the mass and rotation limits with a Salpeter IMF has resulted in a distribution of energies, and by continuing to combine this with models of the formation rates of Pop III GRBs \citep{desouza2011} we find the corresponding distributions of energy and redshift.

As one would expect, the detection duration increases with energy and decreases with redshift. The effect of redshift time dilation to increase the duration is counteracted by the increased luminosity distance reducing the observed flux density, so that the flux density limit is reached earlier. Further, the duration we present the isotropic duration, starting at the jet break time. Since the observed time of the jet break is also affected by redshift time dilation, the total effect on the duration is reduced. Within the range of plausible parameters, we find that the afterglow is always detectable after the jet break time in the SPICA 20 - 24\micron$\,$ band. This implies the JWST and SPICA instruments will be sensitive to the complete intrinsic rate of orphan Pop III GRB afterglows, not only the 0.05\% which are beamed toward them.

The GRB rate from Pop III.2 stars is 1.45$\times 10^2$ yr$^{-1}$, or one every 2.5 days. The energies of Pop III.2 GRBs can create afterglows that remain visible for more than three years and never less than 55 days after the jet break time, far in excess of the interval between bursts. Combining the post jet break duration with the event rates, the all sky number density of Pop III.2 GRBs is 1.76$\times 10^{2}$/4$\pi$ str.

When combined with the fields of view of the proposed telescopes, the number density is 2.78$\times 10^{-6}$ per JWST FOV, implying on average 1 afterglow every $\sim$ 360,000 images; and 2.96$\times 10^{-5}$ per SPICA FOV, or an average of 1 afterglow every $\sim$ 34,000 images.

With these assumptions, we estimate the rate at which Pop III.2 GRBs will align with the EXIST instrument. Assuming an average opening angle of 0.1 rad, the GRBs will be beamed into $\sim 0.05\%$ of the sky. Thus Pop III GRBs will align with EXIST's FOV (1.92 str \citep{grindlay2009}) at a rate of $1.11\times 10^{-1}$yr$^{-1}$, or once every 9 years. This is not taking into account the luminosity of the GRB in the EXIST high energy telescope (HET) observing band or the instrument's sensitivity; this is simply the alignment rate.

\section{DISCUSSION}
JWST and SPICA could potentially detect every Pop III GRB afterglow during isotropic emission. This in part due to the increase in sensitivity of these instruments over their predecessors, and partly due to the high energies of the GRBs. This implies this population of GRBs will be easier to detect from their orphan afterglows than by their prompt gamma-ray emission. It is not a deficiency on the part of any GRB detector, simply a fact that the beaming of GRBs reduces the number of observable events by more than two orders of magnitude.

We now consider the contribution of very massive Pop III.1 stars to the GRB rate. Estimating that a Pop III.1 star would need $>900 M_{\odot}$ (based on the $915M_{\odot}$ progenitor simulated in \citet{suwa2011}) to produce a GRB, we can estimate the Pop III.1 GRB formation rate. We do this by modifying the Pop III.1 GRB formation rate of \citet{desouza2011} in the same way that we modified their Pop III.2 GRB formation rate. The ratio of the formation regions $\left( \int\limits_{900}^{1000} \phi (m) \text{ d}m\right) $ to $\left( \int\limits_{100}^{1000} \phi (m) \text{ d}m\right) $, where $\phi (m)$ is the Salpeter mass function, is 1.15\%. Using this ratio, the formation rate of very massive GRB progenitors integrated over all redshifts is $1.52\times 10^{-4}$ yr$^{-1}$. We have calculated that under the most generous plausible parameters of energy and redshift, the longest duration of a Pop III.1 GRB afterglow is 11.7 years. This extreme duration is still a small fraction of the average $\sim 6.5 \times 10^3$ years between bursts, so the contribution of Pop III.1 stars to the afterglow number density is negligible.

In calculating the number density, we presumed a number of fixed fiducial parameters in order to simulate the GRB afterglows. By performing a few rough calculations, we have found how the number density is affected by changes in each of these parameters (see Table \ref{parameter table}).

\begin{table}
\caption{Number Density Dependence on Model Parameters\label{parameter table}}
\begin{tabular*}{\columnwidth}{@{}r@{} @{}l@{\hspace{1cm}} @{\hspace{1cm}}r @{}l@{}}
\hline \hline
\multicolumn{2}{@{\hspace{-1cm}}c}{Model Parameter}&\multicolumn{2}{@{\hspace{-1cm}}c}{Number Density}\\
\multicolumn{2}{@{\hspace{-1cm}}c}{Range}&\multicolumn{2}{@{\hspace{-1cm}}c}{Dependence}\\
\hline
$0.06 \leq {}$ & $\theta_j \leq 0.14$ & $\propto {}$ & $\theta_j^{1.76}$\\
$10^{-0.4} \leq {}$ & $n \leq 10^{0.4}$ & $\propto {}$ & $n^{-0.13}$ \\
$2.2 \leq {}$ & $p \leq 2.5$ & $\propto {}$ & $p^{-4.34}$ \\
$10^{-1.2} \leq {}$ & $\epsilon_e \leq 10^{-0.8}$ & $\propto {}$ & $\epsilon_e^{-0.27}$ \\
$10^{-2.5} \leq {}$ & $\epsilon_B \leq 10^{-1.5}$ & $\propto {}$ & $\epsilon_B^{-0.15}$\\
\hline
\end{tabular*}
\end{table}

The relations between number density and the parameters $\theta_j$ and n were consistent power laws over a wide range of values. We attribute this to the relative influence of these two parameters on the jet break time. The other parameters showed a more complicated relationship with the number density. The dependence on the electron energy index p peaked near p = 2.2, above which the function had a power law decay with an index of -4.3. $\epsilon_e$ and $\epsilon_B$ had number density minima near the fiducial values, about which the number density was only weakly dependent on the energy fractions; outside of the limits specified in the table for these parameters, the number densities were seen to increase. These are only preliminary findings, as we have yet to investigate the parameter dependence in comprehensive depth or width, or examined the effect of varying multiple parameters at once.

At an all-sky number density of 176, the afterglows will be very difficult to detect, especially since this number represents the unbeamed contribution.

In order to confirm an afterglow, one has to be able to see both the power law spectrum (flux density increase with wavelength) and the temporal decay. Identifying a candidate orphan afterglow requires two observations, while confirming a temporal decay requires at least three observations of the object. The number density of 176/4$\pi$ str equates to $7.14 \times 10^{-2}$ str/afterglow, and we estimate the time it would take for SPICA and JWST to cover this area. The sensitivity limits we used are for an exposure time of 300s, and we assume a total imaging time of ten minutes.

The SPICA instrument will have an FOV of 5' by 5', or $2.11\times 10^{-6}$ str, implying $\sim 34,000$ images/afterglow. SPICA could obtain these images in 235 days. Analysis of the Pop III.2 GRB afterglow durations and rates with respect to energy and redshift, we find that $>90\%$ have detectable durations $>235$ days. This means that imaging the same $7.14 \times 10^{-2}$ str of sky twice with SPICA should identify at least one candidate orphan Pop III GRB afterglow. The orphan afterglow should still being bright enough to confirm as such by follow-up observations.

JWST will have an FOV of 1'.25 by 1'.88, or $1.99\times 10^{-7}$ str. Thus JWST would have to take $\sim 360,000$ images, which would take $\sim 2500$ days. This is longer than the planned minimum lifetime of the telescope (5 years), and far longer than any plausible Pop III GRB afterglow. Within its planned lifetime the telescope could cover $5.23 \times 10^{-2}$ str, and so there would be a $\sim 73\%$ chance of imaging an afterglow. Since the area has to be imaged twice to show an object's temporal decay, the probability that JWST could identify an orphan Pop III GRB afterglow becomes $\sim 37\%$.

Any detection of a high redshift (z $\gtrsim$ 6) orphan afterglow would imply a GRB energy budget at least as high as the most energetic Pop I/II GRBs. While the mass ranges of GRB progenitors is similar across stellar populations (10 - 100 M$_{\odot}$), we expect the Pop III IMF to be broader and/or more top heavy in that range, leading to a higher proportion of higher mass stars leading to higher energy GRBs. Further, the minimum energy expected from a Pop III GRB is near the maximum energy recorded for any GRB, implying that these stars are more efficient at converting mass to GRB energy. Therefore, any GRB of $z \gtrsim 6$ and $E_{iso} \gtrsim 8 \times 10^{54}$ erg would potentially have a Pop III progenitor.

To confirm a Pop III progenitor requires that the spectrum of the afterglow has features consistent with the medium around a Pop III star. Wang et al. (2012) showed that the absorption features of a medium enriched via Pop III supernovae would be easily detectable in the otherwise smooth spectrum of a GRB afterglow. That work considered using the JWST NIRSpec instrument with an exposure time of $10^5$s and R = 1000. We have preliminary findings that the SPICA MCS low resolution spectrograph (LRS, R = 50) will, with a 1 hour exposure time, be capable of detecting Pop III GRB afterglows post-jet break. Further, the flux density sensitivity in this case is not as good as NIRSpec with a $10^5$s exposure. Therefore, if a high redshift orphan GRB afterglow could be identified photometrically, then it should be possible to use JWST or SPICA, with sufficient exposure times ($>10^4$s, possibly $>10^5$s), to obtain R $\sim$ 1000 spectra of the afterglow to determine if its environment is consistent with Pop III stars.

Under a more realistic observing scenario, where deep imaging is performed, the amount of sky covered will be much less. This will reduce the potential number of detections, as in the case of these orphan afterglows it is preferable to search wider than deeper. This is a consequence of the flux density limit being inversely proportional to $\sqrt{\text{exposure time}}$, and the detectable number density being nearly inversely proportional to the flux density limit. While imaging ten times the sky area increases the detection probability tenfold, we have determined from basic calculations that a tenfold increase in exposure time for a given area only increases the detection probability by a factor of $\sim$3.3.

The alternate utility of these space based IR detectors in the study of Pop III GRBs would be as late time afterglow follow-up. This places the burden of initial detection on the GRB detector. By the time JWST and SPICA are operational we expect the primary GRB detector will be EXIST. As determined above, the expected rate of EXIST triggers from Pop III GRBs is one every nine years. The type of bursts which would initiate follow-up with JWST or SPICA would have to be very energetic, with an afterglow indicative of high redshift that appears to decay slowly. Any GRB with a redshift greater than 6 is a candidate for having a Pop III progenitor; in the cases where it is not, the afterglow will not likely still be visible by the time one of these space telescopes can investigate it. Detection of an afterglow several weeks after a high redshift GRB would be indicative of the large energy budget expected of Pop III progenitors.

The number densities, and hence detection probabilities we calculate here are lower limits, based on the most pessimistic models of Pop III star and GRB formation rates and energy distributions. This includes using a bottom-heavy IMF, high metal enrichment, low star and GRB formation efficiencies, and an unweighted rotation distribution. We have also been conservative in calculating the detectable durations of the IR afterglows to JWST and SPICA, by considering a brief exposure time one with of the less sensitive wavelength bands. On the other hand, we have not incorporated how the observable number density is affected by other objects, either by obscuration or gravitational lensing. The number density is simply the average number above a certain brightness threshold at a given time within a given sky area. In the event that the actual Pop III GRB formation rate is greater than the considered most pessimistic case, the number density and thus the chances of observing orphan Pop III GRB afterglows will be higher than what we have calculated here.

\section{ACKNOWLEDGEMENTS}
David Coward is supported by an Australian Research Council Future Fellowship

\clearpage

\end{document}